\documentclass[12pt,floatfix,aps,nofootinbib,superscriptaddress]{revtex4} 
\pdfoutput=1 
\usepackage{amssymb,amsfonts,multirow}

\usepackage{epsfig}
\usepackage{bbm}

\usepackage{placeins}
\usepackage{xspace}
\usepackage{cancel} 
\usepackage{appendix}

\usepackage{amssymb,url}
\usepackage{graphicx}
\usepackage{hyperref}

\usepackage{color}

\usepackage{array}
\usepackage{amsmath}
\usepackage{slashed}

\definecolor{rossoCP3}{cmyk}{0,.88,.77,.40}

\long\def\del #1 \enddel { }

\usepackage{graphicx}
\usepackage{amsmath}
\usepackage{amssymb}
\usepackage{subfigure}

\usepackage{epsfig}

\usepackage{graphicx}
\usepackage{subfigure}
\usepackage{hyperref}

\def\beq{\begin{equation}}
\def\eeq{\end{equation}}

\def\bea{\arraycolsep .1em \begin{eqnarray}}
\def\eea{\end{eqnarray}}
\def\Tr{{\rm Tr}}

\def\eps{\epsilon}

\def\al#1{\alpha_{#1}}

\def\s0#1#2{\mbox{\small{$ \frac{#1}{#2} $}}}
\def\0#1#2{\frac{#1}{#2}}

\def\grgl{\:\hbox to -0.2pt{\lower2.5pt\hbox{$\sim$}\hss}{\raise3pt\hbox{$>$}}\:}
\def\klgl{\:\hbox to -0.2pt{\lower2.5pt\hbox{$\sim$}\hss}{\raise3pt\hbox{$<$}}\:}

\def\lsim{\mathrel{\rlap{\lower4pt\hbox{\hskip1pt$\sim$}}
    \raise1pt\hbox{$<$}}}                
\def\gsim{\mathrel{\rlap{\lower4pt\hbox{\hskip1pt$\sim$}}
    \raise1pt\hbox{$>$}}}                

\newcommand{\ea}[1]{
\begin{align}
#1
\end{align}
}

\newcommand{\seal}[2]{
\begin{subequations}
\label{#1}
\begin{align}
#2
\end{align}
\end{subequations}
}

\begin{document}
${}$\vskip1cm

\title{Radiative symmetry breaking from interacting UV fixed points}
\author{Steven Abel}
\email{s.a.abel@durham.ac.uk}
\affiliation{\mbox{IPPP,
Durham University, South Road, Durham, DH1 3LE}}
\affiliation{\mbox{Theory Division, Cern, Route de Meyrin, 1211 Geneva 23, Switzerland}}
\author{Francesco~Sannino$^{2,\,}$}
\email{sannino@cp3-origins.net}
\affiliation{{\color{rossoCP3}CP${}^3$-Origins} \& the Danish Institute for Advanced Study,
Univ. of Southern Denmark, Campusvej 55, DK-5230 Odense}

\begin{abstract}
\vskip2cm
\noindent
It is shown that the addition of positive mass-squared terms to asymptotically safe gauge-Yukawa theories with perturbative UV fixed points leads to calculable radiative symmetry breaking in the IR. 
This phenomenon, and the multiplicative running of the operators that lies behind it, is akin to the radiative symmetry breaking that occurs in the Supersymmetric Standard Model.   

\vskip8.7cm
{\noindent \footnotesize Preprint:  CERN-TH-2017-066, CP3-Origins-2017-011 DNRF90, IPPP-2017/23}

\end{abstract}
\maketitle
\newpage

\section{Introduction}

The triviality problem and the hierarchy problem are two long-standing and somewhat related issues in field theories containing scalars. 
The former says that introducing an interacting scalar such as the Higgs into the theory results in Landau poles, indicating ultra-violet (UV) incompleteness. 
The latter says that the theory is likely to respond to a UV completion by developing unwelcome relevant operators, in particular mass-terms, that are 
determined by the highest scales in the physics. Traditionally these two problems are addressed independently, but it is interesting to ask if there is a way to solve them simultaneously. 

To see how this might be achieved one can draw some lessons from quark masses, which do not suffer from either problem. There are two reasons why. The first is chiral symmetry which ensures that the quark masses are only multiplicatively renormalised. However by itself this would not be sufficient: QCD is in addition asymptotically free so that there is no triviality problem either, and no UV completion is required. One may be added if desired for the rest of theory but it is entirely reasonable to assume that it does not contribute to chiral symmetry breaking, which enjoys a quite independent status
\footnote{One can contrast this with the supersymmetric Standard Model; there scalar masses are protected by supersymmetry and are also multiplicatively renormalised. The equivalent of chiral symmetry breaking is embodied in the Higgs PQ symmetry breaking $\mu$-term. However successful phenomenology requires a UV completion that has to contribute additional Higgs mass terms. As the Higgs masses are only partially governed by the breaking of PQ symmetry, they cannot be very easily isolated from the rest of the theory like quark masses can. (Practically this manifests itself as an accompanying naturalness problem, namely the so-called $\mu$-problem.)}.

It is evident from this example that the presence of a fixed point in the UV, even a Gaussian one, is a potential solution to both of these problems. When it comes to the hierarchy problem one would say that the symmetry protecting the scalar masses, and playing the same role as chiral symmetry for quark masses, would be quantum conformal symmetry\footnote{We will for this discussion assume that 
scale invariance implies conformality.}, while there would be no triviality problem by construction. However finding calculable and predictive examples of such theories has in the past proven difficult.

 This is because (to use exact RG parlance) the theory flows  along an exactly renormalisable trajectory that traces back precisely to the fixed point.  If it is Gaussian and there are several couplings, then  typically the theory can emerge from the fixed point in multiple directions: virtually \emph{any} trajectory could be an exactly renormalisable one. It is then difficult to determine how the flow out of the UV fixed point constrains the infra-red (IR) physics. What is required for maximum predictivity is for there to be only a single exactly renormalisable trajectory emanating from the fixed point. In the presence of multiple couplings this almost certainly requires that the fixed point be interacting, which in turn makes it hard to find calculable examples.  
 The purpose of this paper is to examine in this context the perturbative example of such a theory presented in ref.\cite{Litim:2014uca}, and developed in refs.\cite{Antipin:2013pya,LMS,Bajc:2016efj,Pelaggi:2017wzr}.

 Our central point will be that, not only does the interacting UV fixed point of this theory provide a UV completion and 
 solve the triviality problem, but it also allows radiative symmetry breaking driven by arbitrary mass-squared operators. The couplings (i.e. the ensemble of masses-squared) are technically natural within the theory in the sense that they run (as an ensemble) multiplicatively from the UV fixed point. In this sense their RG behaviour is similar to, and as controllable as, that of the soft-terms in supersymmetry (SUSY). Moreover we find that, as in the Minimal Supersymmetric Standard Model (MSSM), a positive  mass-squared operator in the UV induces negative mass-squared operators in the IR due to large couplings. 
 
The end result is a calculable radiatively induced symmetry breaking, exactly analogous to that in the MSSM \cite{Ibanez:1982fr}, that is proportional to the explicit degree of flavour breaking in the mass-squared terms. It is a very different and more complete paradigm for radiative breaking than the one normally invoked in the context of scale invariance, namely the Coleman-Weinberg (CW) mechanism \cite{Salam,Coleman:1973jx,gildener,bardeen,scaleinv1,scaleinv2,scaleinv3,Goldberger:2007zk,scaleinv4,scaleinv6,scaleinv7,scaleinv8,scaleinv9,scaleinv10,Coriano:2012nm,scaleinv11,scaleinv115,scaleinv12,scaleinv13,scaleinv14,scaleinv15,scaleinv17,scaleinv18,scaleinv20,scaleinv21,scaleinv22,late1,late2,late3,late4,late5,late9,late10,late11,late12,late13,late14,late15,late16,late17,late18}.
The latter sets the masses to zero at the origin of field space, with some hopeful words that this could well be a prediction of scale invariance. Here we emphasise that UV scale invariance does not 
  prefer {\it any} value for mass-terms, since they are relevant operators, and the fixed point is completely blind to them. 
  Therefore one may perfectly consistently choose a mass-squared parameter to be ``small'' (relative to the dimensional transmutation scale, say) 
 in which case the CW version of radiative symmetry breaking can in principle still operate (although not as it turns out perturbatively in the case we discuss), or one may 
 choose it to be large. Either possibility is consistent with exact quantum UV scale invariance since 
 the ``starting values'' of the dimensionful (i.e. relevant) parameters at some RG scale are free parameters, again much like the quark masses in QCD. 
 Furthermore both cases should be thought of as radiative symmetry breaking, just driven by different operators. 
This picture is of course entirely different from flows governed by IR fixed points, in which relevant operators {\it do} determine the fixed point.

We should explain why an {\it interacting} fixed point  increases the calculability and predictivity. Suppose for example that one wished to compute perturbative corrections to the dimensionless couplings of the effective theory. Such corrections would have a UV ``divergence'' going as $1/\gamma$ where $\gamma$ is the anomalous dimension of the operator at the fixed point. If the fixed point is Gaussian then the  anomalous dimensions at the fixed point are zero and this corresponds to a real divergence which tells us that generically the couplings are simply given by their settings at renormalisation time $-t=\log(\mu_0/\mu)=0$. 

By contrast, if the fixed point is interacting then anomalous dimensions at the fixed point are non-zero, and radiative corrections will simply be finite terms going as $1/\gamma$. In particular they  are insensitive to $\mu_0$ which we may as well take to be infinite. (These points were discussed in some detail in ref.\cite{Abel:2013mya}.)
If the fixed point is interacting and strongly coupled, a perturbative treatment is impossible, but nevertheless non-zero anomalous dimensions make the corresponding couplings insensitive to the precise details of the approach to the UV fixed point, thereby restoring predictivity. Of course the anomalous dimensions may not be large enough to regulate  classically relevant operators: as mentioned above, such operators simply experience multiplicative RG running in the usual way from values chosen at some initial scale $\mu_0$ (much like quark masses), but cannot disrupt the UV fixed point, so the asympotic safety of the set-up is immune to them.

The asymptotically safe theories of ref.\cite{Litim:2014uca} that we will be using here lie somewhere between these two extremes. By choosing a theory with a \emph{weakly} interacting UV fixed point we recover the benefits of predictivity and control over the effective potential, but at the same time keep the theory under good perturbative control. This optimisation is reminiscent of the Banks-Zaks IR fixed point \cite{BZ}, which can be made arbitrarily weakly interacting and hence perturbatively tractable, in a particular (Veneziano)  large-colour/large-flavour limit.

Of course this work follows on from a large body of literature that has discussed  asymptotic safety and more generally the consequences of UV scale invariance 
both with and without gravity:  \cite{asymp-safety,Martin:2000cr,Gies:2003dp, shapo, Gies:2009sv, Braun:2010tt, Wetterich:2011aa, Abel:2013mya, Gies:2013pma, Bazzocchi:2011vr,late7}). (For a review see \cite{Litim:2011cp}).
The object of this paper is to place radiative symmetry breaking in such frameworks on the same footing as it is in the MSSM. 

\section{The Theory, UV Fixed Point and Critical Curve}
\label{Theory}
We begin by describing the behaviour of the weakly interacting gauge-Yukawa theories that we will be using, and in particular their phase diagrams and RG flow. Consider a theory with $SU(N_C)$ gauge fields $A^a_\mu$ and field strength $F^a_{\mu\nu}$ $(a=1,\cdots, N_C)$, $N_F$ flavours of fermions $Q_i$ $(i=1,\cdots,N_F)$ in the fundamental representation, and an $N_F\times N_F$ complex matrix scalar field $H$ uncharged under the gauge group. At the fundamental level the Lagrangian is  $L=L_{\rm YM}+L_F+L_Y+L_H+L_U+L_V$, with
\bea
\label{F2}
L_{\rm YM}&=& - \frac{1}{2} \Tr \,F^{\mu \nu} F_{\mu \nu} + \Tr\left(
\overline{Q}\,  i\slashed{D}\, Q \right) + 
y\,\Tr\left( \overline{Q}\, H\, Q  \right) + \Tr\,(\partial_\mu H ^\dagger\, \partial^\mu H) \nonumber \\
\label{U}
&&-u\,\Tr\,[(H ^\dagger H )^2] 
-v\,(\Tr\,[H ^\dagger H ])^2  \,,
\eea
where $\Tr$ indicates the trace over both color and flavor indices.
The model has four coupling constants given by the gauge coupling, the Yukawa coupling $y$, and the quartic scalar couplings $u$ and the double-trace scalar coupling $v$: 
\beq\label{couplings}
\al g=\frac{g^2\,N_C}{(4\pi)^2}\,,\quad
\al y=\frac{y^{2}\,N_C}{(4\pi)^2}\,,\quad
\al h=\frac{{u}\,N_F}{(4\pi)^2}\,,\quad
\al v=\frac{{v}\,N^2_F}{(4\pi)^2}\,.
\eeq
We have already re-scaled the coupling constants by the appropriate powers of $N_C$ and $N_F$ to work in the Veneziano limit. When necessary we will use a shorthand notation $\alpha_i$ with $i=(g,y,h,v)$.
As mentioned in the Introduction we will be considering the large colour and large flavour Veneziano limit, in order to have an interacting fixed point which is nevertheless arbitrarily weakly coupled. Therefore it is convenient to introduce a control parameter which in the Veneziano limit is a continuous and arbitrarily small constant
 \begin{equation}\label{eps}
\eps=\frac{N_F}{N_C}-\frac{11}{2}\,.
\end{equation}
Asymptotic freedom is lost for positive values of $\eps$.

Ref.\cite{Litim:2014uca} discovered a number of fixed points for this model. However there is one fixed point that is unique in that it has only one relevant direction with the other three being irrelevant. Since every relevant direction loses predictivity (as it is formally zero at the fixed point and must be set by hand) this fixed point is of great interest. To the maximum currently achievable order in perturbation theory and properly respecting the Weyl consistency conditions it is obtained for 
\beq\label{NNLOseries}
\begin{array}{rcl}
\alpha_g^*&=&\; \; \,0.4561\,\eps+0.7808 \,\eps^2
+{\cal O}(\eps^3)\\[.5ex]
\alpha_y^*&=&\; \; \,0.2105\,\eps+0.5082\,\eps^2
+{\cal O}(\eps^3)\\[.5ex]
\alpha_h^*&=&\,  \; \;0.1998\,\eps+0.5042\,\eps^2
+{\cal O}(\eps^3)\,,
\end{array}
\eeq
with the leading coefficients of $\epsilon$ corresponding to $\alpha_g^* = \frac{26}{57}\epsilon +\ldots$, $\alpha_y^* = \frac{4}{19}\epsilon +\ldots$ and $\alpha_h^* = \frac{\sqrt{23}-1}{19}\epsilon +\ldots$ respectively. Note that the quartic scalar self-coupling is essential for this fixed point to exist. The remaining double-trace scalar coupling $v$ has two possible fixed points, one of which is more perturbatively reliable and adds an irrelevant scaling direction to the theory, found to be at
\bea\label{v1v2}
\label{v1}
\al {v1} ^*&=&
\frac{-6 \sqrt{23+4 \epsilon} + 3\sqrt{4 \epsilon +6 \sqrt{23+4 \epsilon}+20}}{4 \epsilon +26}\al g^*
+{\cal O}({\al g^*}^2)\, .
\eea
Numerically $\al {v1} ^*=-0.1373\,\eps$ up to quadratic corrections in $\eps$.

\begin{figure}
\noindent \begin{centering}
\includegraphics[bb=0bp 0bp 220bp 230bp,clip]{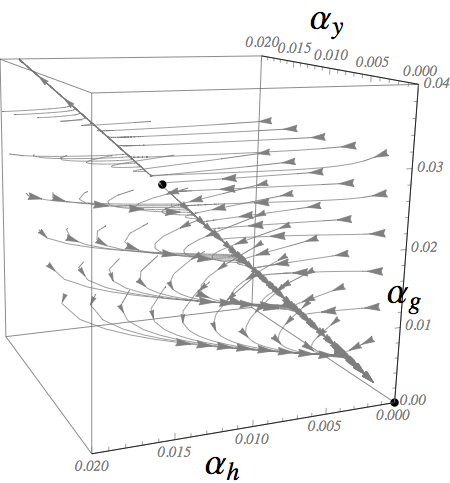}
\par\end{centering}
\protect\caption{The renormalisation group flow of the marginal couplings from the UV fixed point and around the critical
curve, towards the Gaussian IR fixed point.\label{fig:flow}}
\end{figure}

In the presence of more than one relevant direction the flow from the UV would be expected to emanate from a 
critical surface, however with only one relevant direction the flow is along the critical curve shown in Fig.\ref{fig:flow} towards the
IR stable Gaussian fixed point in the infra-red, and is therefore completely determined in terms of a single parameter which could be taken to be the gauge coupling itself. The arrows in the figure are at equal separation in renormalisation ``time'', so it is clear that the flow to the critical curve happens much more rapidly than flow along it. In fact as discussed in Ref.~\cite{Litim:2014uca}  the relative rate of flow is proportional to $\epsilon$. Of course for the present discussion the flow emanates precisely from the UV fixed point of Eq.(\ref{NNLOseries}) marked in black, along the critical curve towards the Gaussian IR fixed point.

In scalar field theories we must also determine if the potential is stable. Ignoring the possible presence of relevant operators for the moment, we see that this is indeed the case at leading order since $\al h^*+\al {v1} ^*>0$, and it also the case for loop corrections as well \cite{Litim:2014uca,LMS}. Therefore there is no Coleman-Weinberg type instability in these models, as will be shown explicitly later in certain directions in field space. Thus the $\alpha^{\ast}_{v1}$ perturbative fixed point is classically viable and becomes increasingly flat in the Veneziano limit, and moreover in the absence of relevant operators the flow never leaves the critical curve.
  
Having identified all the critical coupling values and the scaling dimensions  it is possible to parameterize the gauge coupling and hence the entire flow along the critical curve for any values of the renormalisation time $t = \ln \mu/\mu_0$   by \cite{Litim:2011cp,LMS} 
\seal{critsurf}{
\alpha_g(t) &= \al g^* \left (1+W\left [ \exp\left (-\frac{4}{3} \eps \al g^* (t + \kappa) -1\right) \right ] \right )^{-1} \ ,\\
\al y(t) &= \frac{6}{13} \al g(t) \ , \\
\al h (t) & = 
3\frac{\sqrt{23}-1}{26} \al g (t) \ , \\
\al v (t) & = 
\frac{3\sqrt{20 + 6\sqrt{23}}-6\sqrt{23}}{26} \al g (t) \ ,
}
where $W$ is the Lambert $W$-function (a.k.a. the product log defined by $W(z) e^{W(z)} = z$) and 
$\kappa$ is defined by the initial condition,
\begin{equation}
\alpha_{g}(0)=\frac{\alpha_{g}^{*}}{1+W\left[e^{-\frac{4}{3}\epsilon\alpha_{g}^{*}\kappa-1}\right]}\,.
\end{equation}
Perturbation theory is valid for all values of $t$ as long as $\eps$ is small.  

Since we can access all scales through this set of solutions, the initial gauge coupling is the only free parameter distinguishing different physical systems that flow from the UV fixed point, and must be set by hand in accord with the measurement of the coupling at some scale. However, as mentioned above one can simply use the gauge coupling itself to parameterise the flow along the critical curve
linking the UV interacting fixed point to the IR non-interacting one (also known as the separatrix): it is a monotonically increasing function of $\mu$.

\section{Symmetry breaking}

What happens when we add a classically relevant operator to such a system, in particular of course a mass-squared term for the scalar $H$? As described in the Introduction, as long as the operator remains relevant at the quantum level we do not expect it to affect the UV fixed point, and its status will therefore be equivalent  to that of chiral symmetry breaking mass-terms in QCD, in the sense that it is a parameter which is set at the initial RG scale by physical measurement. There is no question of uncontrolled UV sensitivity because we know that the theory is exactly conformal precisely at the UV fixed point (this is of course the central assumption which unlike the CW mechanism is now motivated by a genuine symmetry). On the other hand being a relevant operator it {\it will} divert the flow away from the IR fixed point. In the current context this flow is precisely the seed for radiative symmetry breaking.

\subsection{A simple example} 

There are a number of different relevant operators that one might consider adding to the theory that can contribute to symmetry breaking. 
They are distinguished by whether or not they explictly break the $SU(N_F)_L\times SU(N_F)_R$ flavour symmetry of the theory. To be concrete we will first consider the mass term,
\begin{equation} 
V\supset \frac{m_\phi^2}{4N_F} \, \left(\Tr({H+H^\dagger})\right) ^2 \, ,
\end{equation}
which explicitly breaks the flavour symmetry to the diagonal, $U(N_F)_L\times U(N_F)_R\rightarrow  SU(N_F)_{diag}$ and picks out just the  scalar component of the trace.

Generally, the RG flow will be on a critical surface whose dimensionality is given by the number of relevant operators (plus one), but if this flavour breaking operator is the dominant one, the flow and stability may be analysed in terms of the corresponding normalised Higgs along its direction, 
\ea{
H =\,\frac{ \phi }{\sqrt{2 N_F}} {\mathbbm{1}_{N_F\times N_F} }\, ,
}
where $\phi$ is real.  We will for the moment restrict our attention to only this direction in field space and assume that a negative  $m^2_\phi $ will ultimately be responsible for symmetry breaking -- in the next subsection we will focus on the main point of the paper, which is that a positive $m^2_\phi$ operator radiatively causes instability in other directions. 

First let us deal with the quartic part of the classical potential of the theory, which  along the $\phi$ direction
reads
\ea{
\label{diagonalpotential}
V^{(4)}_{class} ~=~ \frac{4\pi^2}{N_F^2} ( \al h + \al v ) \, \phi^4 \, .
}
Hence we define  the effective quartic coupling,
\begin{align}
\label{lamlam}
{\lambda} & ~=~{32\pi^{2}} \frac{3}{N_{F}^{2}}\left(\alpha_{h}+\alpha_{v}\right)
\, . \end{align}
It is also useful to define  
\begin{align}
\label{kapkap}
{\kappa} & ~=~{32\pi^{2}}\frac{1}{N_{F}^{2}}\left(3\alpha_{h}+\alpha_{v}\right)
\, . \end{align}
In the absence of $m_\phi^2$ the potential is stable at tree-level, and  
one can also confirm the one-loop stability \cite{LMS}. This essentially rules out the CW form of radiative breaking, because  it is not possible perturbatively to take these theories to a limit in which the crucial  $M(\phi)^4 \log M(\phi)^2 $ terms are dominant. Indeed using the results of the Appendix, the entire one-loop potential along the $\phi$ direction is 
\begin{eqnarray}
\label{cwpot}
V&=& \frac{\lambda }{4!} \phi^4 + \frac{m_\phi^2}{2}  \phi^2 + 
\frac{1}{64\pi^2}\left( m_\phi^2 + \frac{\lambda}{2}\phi^2\right)^2 
\left( \log \frac{m_\phi^2 + \frac{\lambda}{2}\phi^2 }{\mu^2} - \frac{3}{2} \right)\, \nonumber \\
&& ~-~    \frac{(4\pi)^2 }{4 N _FN_C} \alpha^2_y \phi^4  \, 
\left( \log \frac{  (4\pi)^2 \alpha_y  \phi^2  }{\sqrt{N_F N_C}\mu^2} - \frac{3}{2} \right) 
 \nonumber \\
&& ~+~ \frac{(N_F^2-1)}{64\pi^2}\left( \frac{\kappa}{2}\phi^2\right)^2 
\left( \log \frac{ \frac{\kappa}{2}\phi^2 }{\mu^2} - \frac{3}{2} \right)+ \frac{N_F^2}{64\pi^2}\left( \frac{\lambda}{6}\phi^2\right)^2 
\left( \log \frac{ \frac{\lambda}{6}\phi^2 }{\mu^2} - \frac{3}{2} \right)\,  \, .
\end{eqnarray}
The crucial aspect of this expression is that the last line, which contains the contributions from all the orthogonal higgs scalars and pseudoscalars that get a mass,  are according to eqs.(\ref{lamlam}) and (\ref{kapkap}), suppressed by order $\alpha_v$ and $\alpha_h$ with respect to the leading term, despite the factor of $N_F^2$.  From one point of view  this is of course desirable since it ensures that the theory remains perturbative, but it also means that these terms are not able to play off against the tree-level term in order to create a minimum (in contrast with the original CW mechanism which without the constraint of having to be on a renormalisable trajectory could freely set $\lambda \sim \alpha_e^2$). It would of course be interesting to find theories where one could (by varying a parameter such as $m^2_\phi$) go continuously to CW radiative symmetry breaking.  

As promised therefore,  symmetry breaking, if it occurs at all, must be driven by the mass-squared. Its   evolution  may be treated in the same way as for any other coupling in  a perturbative theory. 
It is useful for our later treatment of more complicated flavour structure, to have the relevant expressions to hand of the various contributions to  the RG flow. For this reason (and to be careful about signs and establish conventions) let us summarise  the 
general framework for a theory of scalars $\phi$ with generic $\phi^n$ couplings as  
\begin{align}
\lambda^{(n)} & = \frac{\partial^{n}V}{\partial\phi^{n}}\, ,
\end{align}
where of course for the mass-squared we will take $n=2$, so at the risk of confusion $\lambda^{(2)}\equiv m_\phi^2$.
The main equation to solve is the Callan-Symanzik equation for the $n$-point Green's function,
\begin{align}
\label{CZ}
\left(-\frac{\partial}{\partial t}+\bar{\beta}\frac{\partial}{\partial\lambda^{(n)}}-n\bar{\gamma}\right)\lambda_{eff}^{(n)}=0\, ,
\end{align}
where $t=\log(\phi/\mu_0)$, corresponding to invariance under changes in the cut-off $\mu_0$, of the coupling $\lambda_{eff}^{(n)}(\phi/\mu_0)$ that one calculates directly in the effective field theory.

The bars indicate division by $1+\gamma$: as we will work to one-loop for the evolution of the mass-squareds, they will ultimately be dropped. 
For $n=2$ this gives the anomalous dimension as
\begin{align}
\label{ZZ}
\bar{\gamma} & = - \frac{1}{2}\frac{\partial \log Z}{\partial t}\, ,
\end{align}
where the renormalised fields scale as $\phi \rightarrow \sqrt{Z(t)}\,\phi$, hence $Z=\exp (-2\int \bar{\gamma} \, dt )$. 

In order to solve (\ref{CZ}) we identify $\bar\beta$ as the $t$-derivative of a running coupling $\lambda(t)$ which must be found by solving  
\begin{equation}
\label{genb}
\bar{\beta}=\frac{d\lambda^{(n)}(t) }{dt}=\frac{\partial \lambda_{eff}^{(n)}}{\partial t}+n\bar{\gamma}\lambda^{(n)}\, ,
\end{equation}
with the functional form of the RHS being determined by perturbation theory and eq.(\ref{ZZ}).  The solution for $\lambda^{(n)}_{eff}$ is then given in terms of this coupling,  by 
\begin{align} 
\label{zzz}
\lambda^{(n)}_{eff} = \lambda^{(n)}(t) \, Z^{n/2}\, .
\end{align}

In SUSY for example the $t$-derivative of $\lambda_{eff}^{(n)}$ is zero to all orders due to the non-renormalization theorem, and eq.(\ref{zzz}) simply says 
that $\lambda^{(n)}(t) \propto Z^{-n/2}$: the renormalisation of any coupling including masses  is multiplicative (thereby solving the hierarchy problem) since it comes entirely from absorbing wave-function renormalization. 
On the other hand in pure  $\lambda\phi^4 $ theory one has $\gamma=0$ at one-loop  and the renormalization of $\lambda $ is dominated by the 
effective potential. 

In the present context we require the anomalous dimension of $H$ to one-loop: it will be denoted by $\gamma$ and is simply \cite{Antipin:2014mga}
\begin{align}
\label{gamma}
{\gamma} & ~=~ \alpha_{y} \, .
\end{align}
In addition to the field renormalisation piece, there is a contribution to the running from the cross-term in the one-loop potential, of the form 
\begin{equation}
V ~\supset~\frac{m_\phi^{2}}{2}\phi^{2}\left(1+\frac{\lambda t}{16\pi^{2}}\right)\,,
\end{equation}
where $\lambda \equiv \lambda^{(4)}$ is the quartic coupling.  (When we come to discuss radiatively induced breaking later on, this will be the crucial contribution.) As $m_\phi^2$ is the only coupling with classical dimension, there can be no other contributions to the mass-squared terms at one-loop, as is indeed apparent from eq.(\ref{cwpot}). Thus 
to one-loop (and dropping the bars) \begin{equation}
\label{betam}
{\beta}_{m_\phi^{2}}~=~m_\phi^{2}\left(\frac{\lambda}{16\pi^{2}}+2{\gamma}\right)\,,
\end{equation}
and inserting eq.(\ref{lamlam}) gives 
\begin{equation}
\frac{1}{m_\phi^{2}}{\beta}_{m_\phi^{2}}~=~ 2\alpha_{y}+\frac{6}{N_F^2} (\alpha_{v}+\alpha_{h})\,\, .
\end{equation}
One can conclude that in the Veneziano limit the mass-squared renormalization is dominated by the anomalous dimension 
of the fields and the individual cross-terms die away as $1/N_F^2$. Moreover the beta function is always positive indicating that the operator grows (in absolute terms) in the UV but of course always 
remains relevant\footnote{in the technical sense, and hence not relevant  in the colloquial sense.}. 

Substituting the solutions in eq.(\ref{critsurf}) we obtain 
\begin{equation}
\frac{1}{m_\phi^{2}}{\beta}_{m_\phi^{2}}~=~f\alpha_{g}\, ,\label{eq:beta}
\end{equation}
where 
\begin{equation}
f~=~\frac{12}{13}\left[1+ \frac{3}{4N_F^2}\left( \sqrt{20+6\sqrt{23}}-1-\sqrt{23} \right) \right]\,.
\end{equation}
In the Veneziano limit we find $f\approx 0.92$, with the mass-squared growing in the UV as 
\begin{equation}
m_\phi^{2}~\stackrel{UV}{\longrightarrow}~m_\phi(0)^{2}\left(\frac{\mu}{\mu_{0}}\right)^{f\alpha_{g}^{*}}.
\end{equation}
Of course the reason this does not disrupt the fixed point is that for parametrically small $\alpha_g^*\sim \epsilon$ the $m_\phi^2$ coupling grows much more slowly than $\mu^2$ itself.
On the other hand the physical mass \emph{shrinks }in the IR since $\alpha_{g}(t)\rightarrow0$ there.
Indeed integrating eq.(\ref{eq:beta}) gives the solution 
\begin{align}
m_\phi^{2}(t) & ~=~  m_\phi^{2}(0)\exp\left[f\int_{0}^{t}\alpha_{g}dt\right]\nonumber \\
 & ~=~ m_\phi^{2}(0)\,\omega^{-\frac{3f}{4\epsilon}}\, ,
\end{align}
where 
\begin{equation}
\omega~=~\frac{\alpha_{g}^{*}/\alpha_{g}(t)-1}{\alpha_{g}^{*}/\alpha_{g}(0)-1}\,\, .
\end{equation}
We arrive at a purely perturbative description of the evolution of the mass-squared:
\begin{align}
m_\phi^{2}(t) \,\, &=  \,\, m_{*}^{2}\left(\frac{\alpha_{g}^{*}}{\alpha_{g}}-1 \right)^{-\frac{3f}{4\epsilon}} 
 \,\,\,\, \stackrel{IR}{\longrightarrow}~ \,\,m_{*}^{2}\left(\frac{\alpha_{g}}{\alpha_{g}^{*}}\right)^{\frac{3f}{4\epsilon}},\label{eq:qfp}
\end{align}
where the invariant mass-squared parameter is 
\begin{equation}
m^2_{*}=m^2_\phi(0)\left(\alpha_{g}^{*}/\alpha_{g}(0)-1\right)^{\frac{3f}{4\epsilon}}.
\end{equation}
Note that $m^2_{*}$ is independent of the arbitrary energy scale
$\mu_{0}$ corresponding to $t=0$ at which the flow started. Therefore
each $m^2_*$ parameter defines a unique trajectory for $m_\phi^2(t)$, and the totality of possible flows defines  
a two-dimensional critical surface in $(g,y,u,v,m_\phi^2)$-space.  The importance of eq.(\ref{eq:qfp})
is that (in accord with the whole philosophy  of the renormalisation group) one may now dispense with $\mu_{0}$ and describe the flow
entirely in terms of the RG invariants $m_{*}$, $\alpha_{g}^{*}$,
and the running coupling $\alpha_{g}(t)$.  As was the case for the classically dimensionless couplings, 
 its RG flow is faster by a factor of $1/\epsilon$ than that of the gauge coupling. Moreover this expression makes transparent the multiplicative nature of the mass renormalisation, with the conclusion that in order to have spontaneous symmetry breaking {\em along this direction} the parameter $m_*^2$  has to be negative, implying that $m_\phi^2$ is negative for all RG scales. 

One should of course stop the running around the scale of the Higgs
mass which fixes the relevant values of $\alpha_{g}(t)$ and $m_*$ 
for the desired  masses and gauge coupling, both of which would in principle be determined by measurement. 
Note that $\lambda$ also keeps running until the scale of spontaneous symmetry breaking. Therefore the  minimum
of the tree-level improved potential gives a VEV determined as $\lambda\langle\phi(t)\rangle^{2}=-6m_\phi^{2}(t)$ evaluated with the $t$
parameter corresponding to the value of $|m_\phi|$ itself, which in practice means simply using the appropriate value of $\alpha_g$ measured at the scale of the 
physical Higgs mass. 

As one would expect, the spectrum, including that of the quarks, scales as the invariant $m_*$ and is otherwise 
a function only of $\alpha_g$: 
\begin{align}
m_{higgs}^2/|m_*^2| & \,\,=\,\, 2 \left(\frac{\alpha_{g}^{*}}{\alpha_{g}}-1 \right)^{-\frac{3f}{4\epsilon}} \nonumber \\
m_Q^2/|m_*^2| & \,\,=\,\,   \frac{{2N_F/ N_C} }{\sqrt{20+6\sqrt{63}} -(1+\sqrt{23})} \,\,\left(\frac{\alpha_{g}^{*}}{\alpha_{g}}-1 \right)^{-\frac{3f}{4\epsilon}} \, ,
\end{align}
where the first of these is simply the  usual $m_{higgs}^2  =2 |m_\phi^2|$ relation one has for the Higgs mass of the Standard Model. 
There is no colour breaking here because the Higgs 
is a singlet under colour, so the gluons remain massless. However one could imagine also gauging the flavour in which case the 
flavour gauge boson masses would also scale as $m_*$, although of course one has to be careful to preserve the asymptotic safety of the whole construct.  
In summary, $m_*$ provides a tunable parameter that, much like the quark masses, encompasses the breaking of both scale invariance and flavour symmetry in the entire flow. 

As we hinted above, we are in the above analysis implicitly assuming that a negative $m_*^2$ leads to instability in the $\phi$ direction alone, and not along any of the other directions.  The treatment was also naive in that we have neglected the contribution to the potential of the orthogonal scalars. Their masses are all initially explicitly zero so they do not contribute at leading order, 
but they will start to contribute loop-suppressed terms proportional to $1/N_F^2$  upon resumming the logs. We will see this in a more complete treatment below.    

\subsection{Radiative symmetry breaking}

We now begin to extend the discussion to general flavour  breaking and first demonstrate that the gauge-Yukawa model in eq.(\ref{F2}) has an in-built mechanism for radiatively induced spontaneous symmetry breaking, analogous to the familiar mechanism of the MSSM. That is, even if a {\em positive} parameter such as $m^2_\phi$  is introduced into the theory at a high renormalization scale, the couplings generically lead to radiative instability in orthogonal directions in field space, and hence to spontaneous breaking of flavour \cite{Ibanez:1982fr}. Moreover we shall see various quite striking similarities with radiative symmetry breaking in the MSSM. 

 Consider adding to the previous theory a second  dimensionful operator that breaks the flavour further as $SU(N_F)_{diag}\rightarrow  SU(N_F/2)_{diag}\times SU(N_F/2)_{diag}$. We shall discuss the stability for the VEVs of the corresponding (in terms of symmetry breaking) directions in fields space, namely 
\begin{align}
\label{expform}
H & = \frac{\phi }{\sqrt{2 N_F}} {\mathbbm{1}}_{N_F\times N_F} +  \frac{ h }{\sqrt{2 N_F}} \sigma_1\otimes {\mathbbm{1}}_{N_F/2 \times N_F/2} \, ,
\end{align}
where $\sigma_1$ is the usual Pauli matrix. Again the fields $\phi$ and $h$ are  the real components of complex fields normalised as for example $\frac{1}{\sqrt{2}} (\phi+i\eta )$. For the one-loop potential one has to of course include the mass-squareds of both the scalar and pseudo-scalar fields.

We then add the two mass-squareds for the scalar components into the theory as 
\begin{equation}
V^{(2)}_{class} ~=~\frac{m_\phi^2}{2} \phi^2+ \frac{m_h^2}{2} h^2\, ,
\end{equation}
with superscript $(2)$ indicating quadratic terms. As in the previous example, the pseudoscalars (and indeed any of the other fields)  cannot -- initially at least -- contribute to the running of these terms, as they do not themselves have an explicit mass-squared and therefore do not have the requisite cross-term in the one-loop potential. On the other hand, as we are about to see for $h$, the converse is not true: even if a mass-squared such as $m_h^2$ is zero, it gets renormalised by a non-zero $m_\phi^2$.

Substituting the explicit form of the fields 
in eq.(\ref{expform}) 
into eq.(\ref{F2}), the quartic terms for $\phi$ and $h$ are 
\begin{equation}
V^{(4)}_{class} ~=~\frac{\lambda}{4!} (\phi^4+ h^4) + \frac{\kappa}{4} \phi^2 h^2  \, ,
\end{equation}
where $\lambda$ and $\kappa$ are as defined in eqs.(\ref{lamlam},\ref{kapkap}) and the Appendix. One can  alternatively use the expressions in the Appendix to derive the above, noting that, if we call the particular $h$ generator we have chosen here $T^1$, then $d_{11b}=0$ (where $d_{abc}$ is the totally symmetric $SU(N_F)$ tensor). 

To proceed it is useful to define a set of parameters scaled in terms of $\alpha_g$ that will encompass the various contributions to the running:
\begin{align}
f_\gamma &\stackrel{def}{~=~} {2\gamma }\frac{1}{ \alpha_g}   ~=~  12/13 \,,\nonumber \\
f_\lambda &\stackrel{def}{~=~} \frac{\lambda }{16\pi^2} \frac{1}{ \alpha_g}   ~=~  \frac{9}{13N_F^2}\left[\sqrt{20+6\sqrt{23}}-1-\sqrt{23} \right]\,,\nonumber \\
f_\kappa &\stackrel{def}{~=~} \frac{\kappa }{16\pi^2} \frac{1}{ \alpha_g}   ~=~  \frac{3}{13N_F^2}\left[\sqrt{20+6\sqrt{23}}-3+\sqrt{23} \right]\, ,
\end{align}
where $f=f_\gamma+f_\lambda$.
These ratios (which are all positive) hold along the entire flow (regardless of the presence or otherwise of $m^2_\phi$ and $m^2_h$).  
Note that $\gamma $ is dominant in the large $N_F$ limit. While the running here is not entirely
driven by field renormalization it is as we mentioned earlier certainly dominated by it in the Veneziano limit.
  
Following the procedure outlined above and applying eq.(\ref{genb}) for this case we derive two beta functions;
\begin{align}
\label{eq:bebe}
\beta_{m_{h^2} } &~=~ \alpha_g \left(  f m_h^2 + f_\kappa m_\phi^2 \right)   \nonumber \\
\beta_{m_{\phi^2} } &~=~ \alpha_g \left(  f m_\phi^2 + f_\kappa m_h^2 \right)\, .
\end{align}
These can be diagonalised and solved (using also $\alpha_g = \frac{d}{dt} \log w^{- 3/4\epsilon}$) to give
\begin{align}
\label{answer}
m_h^2 (t) & ~=~ \frac{ w^{-\frac{3f}{4\epsilon}} }{2}\left[ m_h^2(0) (w^{\frac{3f_\kappa}{4\epsilon}}+w^{-\frac{3f_\kappa}{4\epsilon}}) - m_\phi^2(0) (w^{\frac{3f_\kappa}{4\epsilon}}-w^{-\frac{3f_\kappa}{4\epsilon}}) \right] 
\nonumber \\
m_\phi^2 (t) &~=~\frac{ w^{-\frac{3f}{4\epsilon}} }{2}\left[ m_\phi^2(0) (w^{\frac{3f_\kappa}{4\epsilon}}+w^{-\frac{3f_\kappa}{4\epsilon}} ) - m_h^2(0) (w^{\frac{3f_\kappa}{4\epsilon}}-w^{-\frac{3f_\kappa}{4\epsilon}} ) \right] \, .
 \end{align}
 Note that the previous result is obtained for $f_\kappa=0$ as expected (although we should emphasise that for any given set of operators the  $\kappa$ cross-terms and hence $f_\kappa$ is completely determined as above).
  More importantly however, a large positive value of $m_\phi^2$ can generate a negative $m_h^2$ at low scales (where $w> 1$) \emph{even if all mass-squareds are positive in the UV}. Conversely a large positive 
  $m_h^2$ causes instability in the $\phi$ direction. 

This phenomenon is purely an effect of the  one-loop potential and is precisely what happens in the MSSM, 
 where thanks to supersymmetry, the large (s)top Yukawa generates an effective $\kappa$ driving the soft mass-squared term for the associated Higgs negative \cite{Ibanez:1982fr}. 
  
  The effect can be made more explicit by noting that in the large $N_F$ limit, $\epsilon \sim 1/N_F$  whereas $f_\kappa \sim 1/N_F^2$. Therefore we can expand as follows;
\begin{align}
\label{eq:supanswer}
m_h^2 (t) & ~=~ { w^{-\frac{3f}{4\epsilon}} }\left[ m_h^2(0)  - m_\phi^2(0) {\frac{3f_\kappa}{4\epsilon}}\log w +\ldots \right] 
\nonumber \\
m_\phi^2 (t) &~=~{ w^{-\frac{3f}{4\epsilon}} }\left[ m_\phi^2(0)  - m_h^2(0) {\frac{3f_\kappa}{4\epsilon}}\log w +\ldots \right]\, .
 \end{align}
 Since the prefactor scales as   ${\frac{3f_\kappa}{4\epsilon}}\sim 1/N_F$ one must in this simple example ensure that $m_h(0)^2\lesssim m_\phi^2(0)/N_F$ in order for the symmetry breaking to be driven radiatively by $m_\phi^2$. This is really a function of the dominance of the 
 field renormalisation in the running. Of course a vanishingly small  $m_h(0)^2$  could always be invoked by flavour symmetry arguments, however in the following section we shall present an example that does not require such an assumption.

 We should emphasize at this point that the solutions above should be considered to be accurate to one loop and leading log. In principle, and as is about to become clear in the following subsection, \emph{all} of the $m_h^2$ pick up a mass-squared in a similar fashion and these then feed back into $m_\phi^2$ when the logs are resummed during the running. In other words the beta functions for the mass-squareds in all of the orthogonal directions also get contributions from $m_\phi^2$. It is reasonable  to consider only the leading contributions above in particular restricted directions in field space, first because these secondary contributions would be suppressed by more factors of $1/N_F^2$, but also because the mass-squareds do not themselves contribute to the fixed point behaviour, but simply accumulate contributions perturbatively during the running. Nevertheless we now proceed to improve on the situation with a proper treatment of the flavour structure.      
 
 \subsection{General solutions and the role of flavour}
 
 We conclude from eq.(\ref{answer}) that adding a large positive mass-squared operator in the UV will {\em generically} lead to a further spontaneous radiative breaking of flavour symmetry in a multitude of orthogonal directions. But as mentioned above, there was nothing particularly special about the direction $h$ in the above analysis, compared to any of the other flavour breaking directions that we could have chosen. 
 Therefore in order to identify the correct vacuum one should in principle consider the entire complement of Higgses in the theory. 
 
 Let us therefore define the general direction in terms of the generators of flavour (replacing the previous $\phi$ and $\eta$ with $h_0$ and $p_0$ for convenience),  
 \begin{align}
H & = \frac{(h_0 +ip_0)}{\sqrt{2 N_F}} {\mathbbm{1}}_{N_F\times N_F} +  { (h_a+ip_a) } T_a \, ,
\end{align}
where $T_a$ with $a=1\ldots N_F^2-1$  labels the adjoint generators of $SU(N_F)_{diag}$ and by convention  $\Tr(T_a T_a)=\frac{1}{2}$. 
The scalar components in the potential are effectively the hermitian component of $H$ whereas the pseudoscalars are the antihermitian component. 

What is the influence of a positive $m_{h_0h_0}^2$ operator in the other $h_a$ directions? The crucial cross-terms in the potential, $V\supset \kappa_a h_0^2 h_a^2$, arise from the $\Tr (H^\dagger H H^\dagger H)$ operator in eq.(\ref{F2}) and as is clear from eq.(\ref{tomtom}) they are all similar in magnitude, and in fact any generators $T^a$ that also have $d_{aab}=0$ receive degenerate mass-squareds. Therefore if for example $m^2_{h_ah_a}(0)=0$ for all the high scale starting values, then all of these directions receive mass-squareds 
\begin{equation}
m_{h_ah_a}^2 \approx  -\frac{m_0^2(0) }{2}  (w^{\frac{3(f_\kappa-f)}{4\epsilon}}-w^{-\frac{3(f_\kappa+f)}{4\epsilon}}) \qquad \forall \, a\, ,
\end{equation} 
where $f_\kappa$ is as before, and where the approximation is that we are neglecting cross-terms between the $h^2_a$'s which give contributions that are suppressed by powers of $w$. Nevertheless we can conclude that  \emph{every flavour breaking  scalar orthogonal to $h_0$ receives a negative mass-squared}. 
 
 It is interesting to turn the question around and ask when is there guaranteed to be {\em no} instability. From  eq.(\ref{answer}), 
degenerate values of mass-squareds remain degenerate at all scales. This suggests that for all the possible directions to remain stable requires complete degeneracy, 
$m_{h_0h_0}^2\equiv m_0^2=m_{h_ah_a}^2$ $\forall a$, which is satisfied if one adds the only mass-squared operator that breaks no flavour symmetry at all, namely
$ \Tr(H^\dagger H)\, .$

Therefore in order to find a genuine solution to the RG equations that one can legitimately resum, one should begin with the RG equations for the most general set of flavour-breaking operators, and seek a  deviation from flavour universality that is isomorphic under renormalisation: it turns out that a simple suitable structure is 
generator diagonal and universal except for a flavour deviation in only the trace components; namely
\begin{equation}
\label{eq:structure}
V_{class}^{(2)} ~=~ m_{0}^{2}\text{Tr}(H^{\dagger}H)+2\Delta^{2}\sum_a \text{Tr}(T_{a}H^{\dagger})\text{Tr}(T_{a}H)\,,
\end{equation}
which gives 
\begin{equation}
\label{ma2}
m_{h_ah_b}^{2}\,=~m_{p_ap_b}^{2} ~=~(m_{0}^{2}+\Delta^{2})\,\delta_{ab}\,,
\end{equation}
for all the scalar and pseudoscalar $SU(N_F)$ directions, and degenerate trace pseudo-scalar and scalar mass-squareds, $m^2_{h_0h_0}=m^2_{p_0p_0}=m_0^2$.

The renormalisation of the mass-squared couplings can be determined as before (at the cost of considerably more tedium). 
The detailed expressions required to build the one-loop potential for the most general case are given in eq.(\ref{omg-masses}).
 Inserting the structure chosen in (\ref{eq:structure}), we find 
 \begin{align}
 \label{flip}
\beta_{m_{0}^{2}} & ~=~\alpha_{g}\left(f_{m_{0}}m_{0}^{2}+f_{\kappa}^{\Delta}\Delta^{2}\right)\,,\nonumber \\
\beta_{\Delta^{2}} & ~=~ \alpha_{g}f_{\Delta}\Delta^{2}\,,
\end{align}
where using the results from eq.(\ref{app:last}) and inserting the solutions from eq.(\ref{critsurf}) we have 
 \begin{align}
\label{fs}
f_{m_{0}} & ~=~ \frac{6}{13}\left[ \sqrt{20+6\sqrt{23}}  \left(1+\frac{1}{N_F^2}\right) - \frac{2\sqrt{23}}{N_F^2}  \right]\, ,\nonumber \\
f_{\kappa}^{\Delta} &~ =~\frac{6}{13}\left(1-\frac{1}{N_F^2}\right)\left[ \sqrt{20+6\sqrt{23}} -2    \right] ,\nonumber \\
f_{\Delta} & ~=~ \frac{6}{13}\left[ 2 +\frac{\sqrt{20+6\sqrt{23}}   -{2\sqrt{23}}}{N_F^2}   \right]\,\,.
\end{align}
Note that $f_\Delta$ is dominated by the field renormalisation, and that $f_{m_0}-f_{\Delta}\approx f_\kappa^\Delta$ up to corrections of order $1/N_F^2$. The crucial aspect of these beta functions is that no degrees of freedom were neglected in their derivation, and this flavour structure remains intact 
throughout the running. 
In addition note that $\beta_\Delta^2 $ is zero in the limit of vanishing $\Delta$; as anticipated, totally flavour symmetric mass-squareds do not lead to radiative symmetry breaking 
as there can be no preferred direction in field space. Finally, in contrast with the simplistic example above, the cross-term in the beta function coefficients does not vanish in the Veneziano limit.

Eq.(\ref{flip}) can be solved for $\Delta^2$  and the combination 
\begin{equation}
 \tilde{m}^2 ~=~{m}_0^2+\nu  {\Delta}^2 \, , 
 \end{equation}  
 where we define  
 \begin{align}
\nu ~=~ \frac{f^\Delta_\kappa}{f_{m_0}-f_{\Delta}} ~=~ 1-\frac{1}{N_F^2} \, .
 \end{align}
 Since $f^\Delta_\kappa>0$ then $f_{m_0}> f_\Delta$. They  have the following  solutions;
\begin{align}
\label{eq:soln}
{\tilde{m}^2} & ~=~ {\tilde{m}}^2(0) \, w^{-\frac{3f_{m_0}}{4\epsilon}} \, ,\nonumber \\
  {{\Delta}}^2 & ~=~   {{\Delta}}^2(0)  \, w^{-\frac{3f_\Delta}{4\epsilon}} \,. 
\end{align}
As for the simple case,  it is now possible to describe the entire flow in terms of RG invariants; that is defining 
\begin{align}
\tilde{m}^2_{*} & ~=~ \tilde{m}^2(0)\left(\alpha_{g}^{*}/\alpha_{g}(0)-1\right)^{\frac{3f_{m_0}}{4\epsilon}} \nonumber\\
{{\Delta}}_{*}^2 & ~=~ {{\Delta}}^2(0) \left(\alpha_{g}^{*}/\alpha_{g}(0)-1\right)^{\frac{3f_\Delta}{4\epsilon}} \, ,
\end{align}
one can write 
\begin{align}
\label{lats}
{\tilde{m}^2} & ~=~ {\tilde{m}}^2_* \, \left(\frac{\alpha_{g}^{*}}{\alpha_{g}}-1 \right)^{-\frac{3f_{m_0}}{4\epsilon}} \nonumber \\
{{\Delta}^2} & ~=~ {{\Delta}_{*}^2} \, \left(\frac{\alpha_{g}^{*}}{\alpha_{g}}-1 \right)^{-\frac{3f_\Delta}{4\epsilon}} \,. 
\end{align}

\begin{figure}
\noindent \begin{centering}
\includegraphics[scale=0.35,bb=0bp 0bp 800bp 460bp,clip]{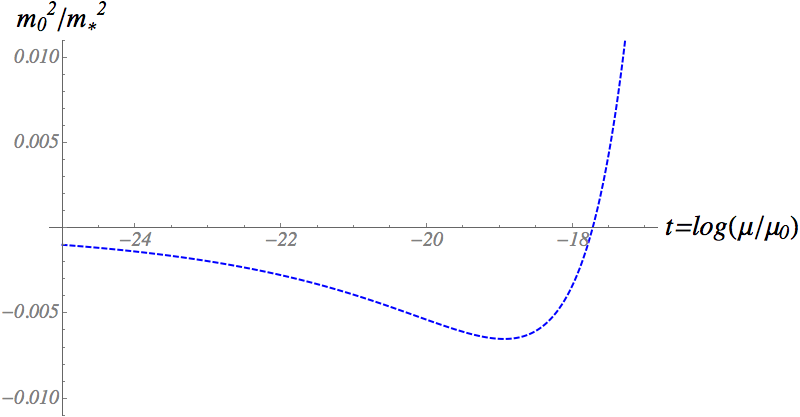}
\par\end{centering}
\protect\caption{A mass-squared that is smaller than the average by 5\% being driven negative radiatively, (where the initial value at $t=0$ is 0.99). We take $\epsilon = 0.1$ in the Veneziano limit ($N_F\rightarrow\infty$). \label{fig:flow2}}  
\end{figure}

With this solution to hand, it is now possible to see how the flavour structure drives radiative symmetry breaking. Consider the case of a slightly positive  $\Delta^2_*$, that is, the $SU(N_F)$ flavour breaking directions 
are given a slightly larger mass-squared than the trace $h_0$ direction. 
According to eq.(\ref{eq:soln}) $\tilde{m}^2$ shrinks very rapidly in the IR as $w^{-\frac{3f_{m_0}}{4\epsilon}} \rightarrow w^{-2.4/\epsilon}$ (recalling that $w$ grows in the IR). On the other hand  the 
deviation ${ \Delta}^2$ also shrinks, but much more slowly, as $w^{-\frac{3f_\Delta}{4\epsilon}}\rightarrow w^{-0.7/\epsilon}$. 
Because $f_{m_0}$ is  greater than $f_\Delta$, the dominance of $\Delta^2$ in the IR is inevitable. Indeed the mass-squareds for the different  components are  
\begin{align}
m_0^2 & ~~=~~ \tilde{m}^2_*  \left(\frac{\alpha_{g}^{*}}{\alpha_{g}}-1 \right)^{-\frac{3f_{m_0}}{4\epsilon}}  -~      {{\Delta}_{*}^2}\, \nu  \, \left(\frac{\alpha_{g}^{*}}{\alpha_{g}}-1 \right)^{-\frac{3f_\Delta}{4\epsilon}}\, , \nonumber \\
m_{a=1\ldots N_F^2-1}^2 & ~=~ \tilde{m}^2_*  \left(\frac{\alpha_{g}^{*}}{\alpha_{g}}-1 \right)^{-\frac{3f_{m_0}}{4\epsilon}}  +~      {{\Delta}_{*}^2}\left( 1 - \nu \right)  \, \left(\frac{\alpha_{g}^{*}}{\alpha_{g}}-1 \right)^{-\frac{3f_\Delta}{4\epsilon}}\, ,
\end{align}\\
with  the $\Delta^2$ piece eventually coming to dominate in the IR. Note that since $1-\nu = 1/N_F^2$, in the large $N_F^2$ limit  light $h_a$ directions are collectively driving a much larger negative mass-squared for the 
single $h_0$ direction.  (The sum of the  mass-squareds is approximately zero). We conclude that a positive $m_0^2$ is driven entirely negative in the IR if we begin with a preponderance of orthogonal slightly heavier directions in the UV. 
An example flow is shown in figure~\ref{fig:flow2}. As is evident from the figure a minimum appears where the deviation  $\Delta^2$ overcomes the running average mass-squared. 

Even if the flavour breaking is tiny (for example the 5\% shown in the figure), this happens very quickly, and  the potential itself develops a minimum at the transmutation scale corresponding to the minimum value of  
$m_0^2$; defining 
\begin{equation}
R_{*}~=~ \frac {{\Delta}_{*}^2}{{{\tilde{m}}}^2_*}\, ,
\end{equation} 
the mass-squared (and hence the potential) forms a minimum at  
\begin{align}
\frac{\alpha_{g}^{*}}{\alpha_{g,min}}-1  &~\approx ~ \left(\frac{f_\Delta}{f_{m_0}} \nu R_{*}\right)^{-{\frac{4\epsilon}{3(f_{m_0}-f_\Delta)}}}  \, ,  \nonumber \\
{m}_{0,min}^2 &~\approx~ - \tilde{m}_*^2\ 
\frac{f_{m_0}-f_{\Delta}}{f_{\Delta}} \left( R_* \nu \frac{f_\Delta}{f_{m_0}} \right) ^{\frac{f_{m_0}}{f_{m_0}-f_{\Delta}}}
\,. \end{align} 
 For the example in figure \ref{fig:flow2}, where $R_{*}=0.05$ and $\epsilon=0.1$, the above approximations give  
  $\alpha_{g,min} = 0.44 \,\alpha_g^*$ and $m_{0,min}^2\approx -6.5\times 10^{-3} \,{{m}}^2_* $.
 Note that for small $\epsilon$ in the Veneziano limit one has 
 \begin{align}
\alpha_{g,min} &~\stackrel{\epsilon\rightarrow 0}{\longrightarrow} ~~ \frac{1}{2} \alpha_g^* \, . 
\end{align}
In other words the minimum forms at precisely the scale where the theory is 
passing from the UV fixed point, and assuming more standard Gaussian IR fixed point behaviour. 
Finally note that if we had chosen negative $\Delta^2$ 
 the reversed pattern of breaking would have occurred, with the trace $h_0$ direction being the only stable and very heavy direction, with a mass-squared balancing order $N_F^2$  
 very small negative mass-squareds for all the orthogonal directions.

\section{Conclusions}

We have studied the stability properties of the class of perturbative UV fixed point theories introduced in  ref.\cite{Litim:2014uca}, in the presence of additional scalar mass-squared terms. 
It is important to realise that such terms, being relevant operators, may take any value in a 
scenario of asymptotic safety without disrupting the fixed point. As such their status is similar to that of the quark masses in QCD: they are simply set by hand at 
some scale and are fully controlled and multiplicatively renormalised along the entire RG trajectory.  Indeed the value of all the relevant operators everywhere along the flow is completely determined by a set of corresponding RG invariants. 

This general picture, in which the trajectories of relevant operators (for example $m^2_*$ in our case) are determined by  a set of tunable  RG invariants that  
defines a particular model, while the marginal operators are all (except for one) determined by a UV fixed point, is a familiar one  in the context of the exact renormalisation 
group. However it is certainly novel to be able to treat it perturbatively. 

Such a treatment reveals that these theories exhibit an interesting form of calculable radiatively induced symmetry breaking, that is exactly analogous to that in the MSSM \cite{Ibanez:1982fr}. 
It was found that a generic set of \emph{positive} but flavour violating mass-squared terms  automatically induce a minimum radiatively,  whose depth is determined by the absolute value of the {\emph {flavour violation}}.  
Moreover the minimum inevitably appears at the precise  scale where the UV fixed point first loses control over the running, and the theory comes under the more familiar  influence of the Gaussian IR fixed point. 
This is a novel radiative symmetry breaking phenomenon that we believe deserves further study.

\appendix

\section{results for potential}

\noindent We collect the results required for the computation of the beta functions. The fields contributing to the one-loop potential 
are decomposed into real and pseudo-scalar flavour
breaking directions as 
\begin{equation}
H=\frac{h_0+ip_{0}}{\sqrt{2N_{F}}}+\left(h_{a}+ip_{a}\right)T^{a}.
\end{equation}
For convenience we use $h_{A}\equiv\{h_0,\,h_{1\ldots N^{2}-1}\}$
with 
\begin{equation}
T^{0}=\frac{1}{\sqrt{2N_{F}}}\,,
\end{equation}
and with small letters running as $a=1\ldots N_{F}^{2}-1$ and capitals
as $A=1\ldots N_{F}^{2}-1$. In principle to determine the correct vacuum one should consider the beta functions of the most general  
 mass-squared terms, 
are, 
\begin{equation}
V^{(2)}_{class}=\frac{1}{2} m_{h_Ah_B}^{2}  h_A h_B + \frac{1}{2} m_{p_Ap_B}^{2}  p_A p_B+m_{p_Ah_B}^{2}  p_A h_B
\, .\end{equation}
Therefore we examine the potential in the background of all the scalars $h_{A}$ and pseudoscalars $p_A$.
Using the identity 
$T_aT_b = \frac{1}{2} \left( \frac{\delta_{ab}}{N_F}+(d_{abk}+if_{abk})T^k \right)$, the relevant pieces are extracted from the tree-level potential which is  
 \begin{eqnarray}
\label{potty}
V_{class}  ~ = ~ \frac{1}{2} m_{h_Ah_B}^{2}  h_A h_B + \frac{1}{2} m_{p_Ap_B}^{2}  p_A p_B+m_{p_Ah_B}^{2}  p_A h_B+\frac{(v+u/N_{F})}{4}(h_{A}^{2}+p_{A}^{2})^{2}~+ \mbox{  \hspace{1.0cm}  }  &&  \\
\frac{u}{8}\mbox{ $ \hspace{-0.1cm}  \left( \hspace{-0.1cm}  (h_{a}h_{b} \hspace{-0.1cm} +\hspace{-0.1cm} p_{a}p_{b})d_{abk}  +2p_{a}h_{b}  f_{abk}+\frac{4 (h_0 h_k+ p_0 p_k)}{\sqrt{2N_F}}  \right)   \hspace{-0.1cm} \left(  \hspace{-0.1cm}  (h_{c}h_{d}\hspace{-0.1cm} +\hspace{-0.1cm} p_{c}p_{d})d_{cdk}  +2p_{c}h_{d}  f_{cdk}+\frac{4(h_0 h_k+ p_0 p_k)}{\sqrt{2N_F} } \right) \,,\nonumber  $} 
\end{eqnarray}
\noindent 
where $d_{abc}$ and $f_{abc}$ are the usual totally symmetric tensor
and antisymmetric structure constants respectively of $SU(N_{F})$.  Note that repeated indices are summed.

The $u$-terms can be significantly simplified by defining 
\begin{align}
\hat{d}_{ABC} & =\begin{cases}
d_{ABC} & A,B,C\neq0\\
\sqrt{\frac{2}{N_{F}}}\delta_{BC} & A=0\,.
\end{cases}\\
\hat{f}_{ABC} & =\begin{cases}
f_{ABC} & A,B,C\neq0\\
0 & A\,\text{or}\,B\,\text{or}\,C=0\,.
\end{cases}
\end{align}
The full tree-level potential then becomes  \begin{eqnarray}
\label{potty}
V_{class}  ~ = ~\frac{1}{2} m_{h_Ah_B}^{2}  h_A h_B + \frac{1}{2} m_{p_Ap_B}^{2}  p_A p_B+m_{p_Ah_B}^{2}  p_A h_B+
\frac{v}{4}(h_{A}^{2}+p_{A}^{2})^{2}~+ \mbox{  \hspace{1.5cm}  }  &&  \\
\frac{u}{8}\mbox{ $ \hspace{-0.1cm}  \left( \hspace{-0.1cm}  (h_{A}h_{B} \hspace{-0.1cm} +\hspace{-0.1cm} p_{A}p_{B})\hat{d}_{ABK}  +2 p_{A}h_{B}  \hat{f}_{ABK} \right) 
  \hspace{-0.1cm} \left(  \hspace{-0.1cm}  (h_{C}h_{D}\hspace{-0.1cm} +\hspace{-0.1cm} p_{C}p_{D})\hat{d}_{CDK}  +2p_{C}h_{D}  \hat{f}_{CDK} \right) \,,\nonumber  $} 
\end{eqnarray}
The field dependent mass-squareds derived from eq.(\ref{potty}) are 
\begin{align}
M_{h_{A}h_{B}}^{2} & \,=\,m_{h_Ah_B}^{2}+v\left(\delta_{AB}\,(h_{C}^{2}+p_C^2)+2\,h_{A}h_{B}\right)\,\nonumber \\
& \qquad + \frac{u}{2} \hat{d}_{ABK} \left( ( h_{C}h_{D}+p_Cp_D) \,\hat{d}_{CDK}+2\, p_C h_D \hat{f}_{CDK} \right) \nonumber \\
& \qquad + {u}\left(  h_{C}\, \hat{d}_{ACK}-p_C  \hat{f}_{ACK} \right) \left(  h_{D} \,\hat{d}_{BDK}-p_D  \hat{f}_{BDK} \right) \, ,
\nonumber \\
M_{p_{A}p_{B}}^{2} & \,=\,m_{p_Ap_B}^{2}+v\left(\delta_{AB}\,(h_{C}^{2}+p_C^2)+2\,p_{A}p_{B}\right)\,\nonumber \\
& \qquad + \frac{u}{2} \hat{d}_{ABK} \left( ( h_{C}h_{D}+p_Cp_D)\, \hat{d}_{CDK}+2 \, p_C h_D \hat{f}_{CDK} \right) \nonumber \\
& \qquad + u \left( p_{C} \hat{d}_{ACK}+h_C  \hat{f}_{ACK} \right) \left(  p_{D} \hat{d}_{BDK}+h_D  \hat{f}_{BDK} \right) \, ,
\nonumber \\
M_{p_{A}h_{B}}^{2} & \,=\,m_{p_Ah_B}^{2}+2v \,p_{A}h_{B} \,\nonumber \\
& \qquad + \frac{u}{2} \hat{f}_{ABK} \left( ( h_{C}h_{D}+p_Cp_D)\, \hat{d}_{CDK}+2 \, p_C h_D \hat{f}_{CDK} \right) \nonumber \\
& \qquad + u \left( p_{C} \hat{d}_{ACK}+h_C  \hat{f}_{ACK} \right)  \left(  h_{D} \,\hat{d}_{BDK}-p_D  \hat{f}_{BDK} \right) \, \,.
\label{omg-masses}\end{align}
Note that when it comes to the renormalization of the mass-squareds,
only those terms with a direct mass-squared can contribute, although
of course all terms  contribute to the quartic coupling
renormalization in the usual way regardless of the flavour breaking.

It is worth
highlighting the generator independence of the above relations:
in a  background of only  real scalars, the terms for $M_{h_{a}h_{a}}^{2}$ and $M_{h_0h_0}^{2}$ can be rewritten 
\begin{align}
\label{tomtom}
M_{h_{a}h_{a}}^{2}\, & ~=~\,m_{a}^{2}+\frac{\lambda}{2}\, h_{a}^{2}+\frac{\kappa}{2}\,  h_{C\neq a}^{2}\,+\, \mbox{\it cross-terms involving  $d_{abc}$} \,,\nonumber \\
M_{h_0h_0}^{2}\, & ~=~ \,m_{h_0}^{2}+\frac{\lambda}{2}\,h_0^{2}+\frac{\kappa}{2}\, h_{a}^{2}\,,
\end{align}
where as in the text, the coefficients are \begin{align} \lambda/2 & ~=~ 3\left(\frac{u}{N_{F}}+v\right)=\frac{3(4\pi)^{2}}{N_{F}^{2}}(\alpha_{h}+\alpha_{v}) \, ,
\nonumber \\
\kappa/2 &~=~ \left(3\frac{u}{N_{F}}+v\right)=\frac{(4\pi)^{2}}{N_{F}^{2}}(3\alpha_{h}+\alpha_{v})\, .\end{align} 

\subsection{The degenerate example}

\noindent Now let us specialise to the specific generator-diagonal structure
considered in the text,
\begin{equation}
V_{class}^{(2)}~=~m_{0}^{2}\text{Tr}(HH^{\dagger})+2\Delta^{2}\delta_{ab}\text{Tr}(HT^{a})\text{Tr}(H^{\dagger}T^{b})\,.\label{eq:vclassapp}
\end{equation}
We will derive $f_{m_{0}^{2}}$ and $f_{\Delta^{2}}$ in the beta
function for $m_{0}^{2}$ and $\Delta^{2}$, defined as 
\begin{align}
\beta_{m_{0}^{2}} & ~=~\alpha_{g}\left(f_{m_{0}}m_{0}^{2}+f_{\kappa}^{\Delta}\Delta^{2}\right)\,,\nonumber \\
\beta_{\Delta^{2}} & ~=~ \alpha_{g}f_{\Delta}\Delta^{2}\,.
\end{align}
Note that we do not expect to find a term proportional to $m_{0}^{2}$
in $\beta_{\Delta^{2}}$ because the completely
flavour-symmetric system should be stable against
radiative corrections to $\Delta$, so its beta function should vanish when $\Delta=0$. Note also that $f_{m_{0}}\alpha_{g}\,\equiv\,\gamma_{\hat{S}}$
can be identified as the anomalous dimensions of the operator $\hat{S}=\text{Tr}(HH^{\dagger})$.
This will give us a useful cross-check with the results of ref.\cite{Antipin:2014mga}. 

Inserting eq.(\ref{eq:vclassapp}) into eq.(\ref{omg-masses}), we find the following
cross-terms contributing to the mass-squareds in the one-loop potential;
\begin{align}
\partial_{t}V^{(m_{h_{a}h_{a}}^{2})}\,\,=\,\, & \frac{\left(m_{0}^{2}+\Delta^{2}\right)}{16\pi^{2}}\sum_{a}\left(v\left[2h_{a}^{2}+\sum_{C}(h_{C}^{2}+p_{C}^{2})\right]+\frac{u}{N_{F}}\left[\sum_{C}(h_{C}^{2}+p_{C}^{2})+2h_{a}^{2}+2h_{0}^{2}\right]\right)\nonumber \\
 & \,\,\,\,\,\,\,\,\,\,\,\,\,\,\,\,\,\,\,\,\,\,\,\,\,\,\,\,\,\,\,\,\,\,\,\,\,\,\,\,\,\,\,\,\,\,\,\,\,\,\,\,\,\,+\frac{\left(m_{0}^{2}+\Delta^{2}\right)}{16\pi^{2}}\sum_{a}u\left[h_{c}h_{d}d_{ack}d_{adk}+p_{c}p_{d}f_{ack}f_{adk}\right]\,,\\
\partial_{t}V^{(m_{p_{a}p_{a}}^{2})}\,\,=\,\, & \frac{\left(m_{0}^{2}+\Delta^{2}\right)}{16\pi^{2}}\sum_{a}\left(v\left[2p_{a}^{2}+\sum_{C}(h_{C}^{2}+p_{C}^{2})\right]+\frac{u}{N_{F}}\left[\sum_{C}(h_{C}^{2}+p_{C}^{2})+2p_{a}^{2}+2p_{0}^{2}\right]\right)\nonumber \\
 & \,\,\,\,\,\,\,\,\,\,\,\,\,\,\,\,\,\,\,\,\,\,\,\,\,\,\,\,\,\,\,\,\,\,\,\,\,\,\,\,\,\,\,\,\,\,\,\,\,\,\,\,\,\,+\frac{\left(m_{0}^{2}+\Delta^{2}\right)}{16\pi^{2}}\sum_{a}u\left[p_{c}p_{d}d_{ack}d_{adk}+h_{c}h_{d}f_{ack}f_{adk}\right]\,,\\
\partial_{t}V^{(m_{h_{0}h_{0}}^{2})}\,\,=\,\, & \frac{m_{0}^{2}}{16\pi^{2}}\left(v\left[2h_{0}^{2}+\sum_{C}(h_{C}^{2}+p_{C}^{2})\right]+\frac{u}{N_{F}}\left[\sum_{C}(3h_{C}^{2}+p_{C}^{2})\right]\right)\,,\\
\partial_{t}V^{(m_{p_{0}p_{0}}^{2})}\,\,=\,\, & \frac{m_{0}^{2}}{16\pi^{2}}\left(v\left[2p_{0}^{2}+\sum_{C}(h_{C}^{2}+p_{C}^{2})\right]+\frac{u}{N_{F}}\left[\sum_{C}(h_{C}^{2}+3p_{C}^{2})\right]\right)\,,
\end{align}
where for example $\partial_{t}V^{(m_{h_{a}h_{a}}^{2})}$ denotes
the terms coming from cross-products with $m_{h_{a}h_{a}}^{2}$. Using the standard $SU(N_{F})$ identities
(repeated indiced summed),
\begin{align}
d_{aak} & =0\,\,;\,\,\,d_{ack}d_{adk}=\frac{N_{F}^{2}-4}{N_{F}}\delta_{cd}\,\,;\nonumber \\
f_{ack}d_{adk} & =0\,\,;\,\,\,f_{ack}f_{adk}=N_{F}\delta_{cd}\,\,,
\end{align}
and summing we find,
\begin{equation}
\partial_{t}V^{(m_{h_{a}h_{a}}^{2})}+\partial_{t}V^{(m_{p_{a}p_{a}}^{2})}=
\mbox{\hspace{-0.1cm} $\frac{\left(m_{0}^{2}+\Delta^{2}\right)}{16\pi^{2}}\left[2v\sum\left(h_{a}^{2}+p_{a}^{2}\right)+2\left(v\hspace{-0.1cm}+\hspace{-0.1cm}\frac{2u}{N_{F}}\right)(N_{F}^{2}-1)\sum(h_{C}^{2}+p_{C}^{2})\right]$},
\end{equation}
where there are some notable cancellations of the structure constant
terms, and 
\begin{equation}
\partial_{t}V^{(m_{h_{0}h_{0}}^{2})}+\partial_{t}V^{(m_{p_{0}p_{0}}^{2})}=\frac{m_{0}^{2}}{16\pi^{2}}\left(4\left(v+\frac{u}{N_{F}}\right)\left[\sum_{C}(h_{C}^{2}+p_{C}^{2})\right]-2v\sum_{a}(h_{a}^{2}+p_{a}^{2})\right)\,.
\end{equation}
In total then, the contributions can be divided into pieces proportional
to $\sum_{C}(h_{C}^{2}+p_{C}^{2})$ that contribute to $\beta_{m_{0}^{2}}$
, and proportional to $\sum_{a}(h_{a}^{2}+p_{a}^{2})$ that contribute
to the running of $\Delta^{2}$:
\begin{align}
\partial_{t}V & \,=\,\frac{m_{0}^{2}}{16\pi^{2}}\left[2v(N_{F}^{2}+1)+4uN_{F}\right]\sum_{C}(h_{C}^{2}+p_{C}^{2})\,
\\
&\qquad  +\,\frac{\Delta^{2}}{16\pi^{2}}\left[2\left(v+\frac{2u}{N_{F}}\right)(N_{F}^{2}-1)\right]\sum_{C}(h_{C}^{2}+p_{C}^{2})
 \,+\,\frac{\Delta^{2}}{16\pi^{2}}2v\sum_{a}\left(h_{a}^{2}+p_{a}^{2}\right)\,.\nonumber
\end{align}
As expected the terms proportional to $m_{0}^{2}\sum_{a}\left(h_{a}^{2}+p_{a}^{2}\right)$ cancel in a non-trivial manner. 

The beta function coefficients are found (as in the main body of the
text) by reading off (twice) the coefficient of the corresponding
term in $\partial_{t}V$, and adding $2\gamma=2\alpha_{y}$ to the
diagonal pieces for the anomalous dimension of the fields. We find
\begin{align}
\label{app:last}
f_{m_{0}}\alpha_{g} & =2\alpha_{y}+4\alpha_{v}\left(1+\frac{1}{N_{F}^{2}}\right)+8\alpha_{h}\, ,\nonumber \\
f_{\kappa}^{\Delta}\alpha_{g} & =\left(4\alpha_{v}+8\alpha_{h}\right)\left(1-\frac{1}{N_{F}^{2}}\right)\, ,\nonumber \\
f_{\Delta}\alpha_{g} & =2\alpha_{y}+\frac{4}{N_F^2}\alpha_{v}\,\,.
\end{align}
As a check we can confirm that the first of these is the anomalous dimension $\gamma_{\hat S}$ of the singlet composite operator, calculated in  \cite{Antipin:2014mga}. For use in the 
text note that 
\begin{align}
\nu ~=~  \frac{f^\Delta_\kappa}{f_{m_0}-f_{\Delta}} ~=~ 1-\frac{1}{N_F^2}\, .
\end{align}

\bibliographystyle{apsrev4-1}

\end{document}